\documentclass[aps,prl,twocolumn,nofootinbib,balance,superscriptaddress,floats]{revtex4}

\usepackage[a4paper, total={7.2in, 10in}]{geometry}
\usepackage{latexsym}
\usepackage{dcolumn}
\usepackage{amsmath}
\usepackage{epsf}
\usepackage{float}
\usepackage{hyperref}

\usepackage[pdftex]{graphicx}
\usepackage{epstopdf}
\usepackage{pdfpages}

\usepackage{soul}

\usepackage{upgreek}
\usepackage{tabularx}

\begin{document}


\title{Integrated RF-photonic Filters via Photonic-Phononic Emit-Receive Operations}

\author{Eric A. Kittlaus$^{1}$}
\noaffiliation
\author{Prashanta Kharel}
\affiliation{Department of Applied Physics, Yale University, New Haven, CT 06520 USA.}
\author{Nils T. Otterstrom}
\affiliation{Department of Applied Physics, Yale University, New Haven, CT 06520 USA.}
\author{Zheng Wang}
\affiliation{Microelectronics Research Center, Department of Electrical and Computer Engineering, The University of Texas at Austin, Austin, TX 78758 USA.}
\author{Peter T. Rakich$^{1}$}
\noaffiliation


\date{\today}

\begin{abstract}

The creation of high-performance narrowband filters is of great interest for many RF-signal processing applications. To this end, numerous schemes for electronic, MEMS-based, and microwave photonic filters have been demonstrated. Filtering schemes based on microwave photonic systems offer superior flexibility and tunability to traditional RF filters. However, these optical-based filters are typically limited to GHz-widths and often have large RF insertion losses, posing challenges for integration into high-fidelity radiofrequency circuits. In this article, we demonstrate a novel type of microwave filter that combines the attractive features of microwave photonic filters with high-Q phononic signal processing using a photonic-phononic emit-receive process. Through this process, a RF signal encoded on a guided optical wave is transduced onto a GHz-frequency acoustic wave, where it may be filtered through shaping of acoustic transfer functions before being re-encoded onto a spatially separate optical probe. This emit-receive functionality, realized in an integrated silicon waveguide, produces MHz-bandwidth band-pass filtering while supporting low RF insertion losses necessary for high dynamic range in a microwave photonic link. We also demonstrate record-high internal efficiency for emit-receive operations of this type, and show that the emit-receive operation is uniquely suitable for the creation of serial filter banks with minimal loss of fidelity. This photonic-phononic emitter-receiver represents a new method for low-distortion signal-processing in an integrated all-silicon device.
\end{abstract}


\maketitle

\section{Introduction}

Rapid growth in wireless communications has created a pressing demand for high-performance and reconfigurable schemes for radio-frequency (RF) signal processing, spectral awareness, and frequency-agile filtering of microwave signals \protect{\cite{Seeds06,marpaung13,urick2015fundamentals}}. As a result, considerable work has been directed toward creating high-performance, frequency-agile RF filters with narrow spectral bandwidths. While traditional electronic filters can support narrow bandwidths, it is challenging to design such filters with rapid wideband tunability \protect{\cite{Capmany2007}}. By contrast, filters based on microwave (or RF-) photonic platforms offer superior tunability and the potential for integration in complex photonic circuits \protect{\cite{capmany05,Capmany2007,marpaung13}}. However, integrated all-optical RF-photonic filters are typically limited to spectral resolutions above 100-1000 MHz and often have trade-offs between performance metrics such as bandwidth, rejection, and nonlinearity \protect{\cite{marpaung13,capmany05,marpaung}}. Moreover, many of these systems are plagued by high RF insertion losses due to inefficiencies in RF-optical conversion and low power handling in photonic integrated circuits \protect{\cite{urick2015fundamentals}}.

On-chip stimulated Brillouin scattering (SBS) has recently emerged as a powerful tool for hybrid photonic-phononic signal processing operations, many of which have no analogue in all-optical systems \protect{\cite{pant2014,marpaung,shinpper,CasasBedoya15}}. SBS, which results from the interaction between optical fields and acoustic phonons in the MHz-GHz range, has been used for optical amplification \protect{\cite{olsson,pant,Kittlaus2016,Kittlaus2017,Morrison17}}, RF filtering \protect{\cite{Tanemura02,Zadok07,Vidal07,zhang12}} and signal processing \protect{\cite{yao98,loayssa,Li2013,Merklein16}}, nonreciprocity \protect{\cite{Huang11,Kang2011,kim2015,Dong2015}} and time delay \protect{\cite{okawachi05,Zhu07}}. The recent realization of SBS in silicon waveguides \protect{\cite{rakichprx,shinnatcomm,vanlaernatphoton}} has been used to create all-silicon amplifiers \protect{\cite{van2015net,Kittlaus2016,Kittlaus2017}}, a silicon-based Brillouin laser oscillator \protect{\cite{Otterstrom2017}}, and new architectures which permit MHz-bandwidth filtering operations \protect{\cite{marpaung,shinpper,CasasBedoya15}} with the potential for large-scale integration into silicon photonic circuits.

A promising method to develop Brillouin-based RF filters that has recently been explored harnesses phonon-mediated emit-and-receive coupling between distinct nanophotonic waveguides \protect{\cite{shinpper}}. Through this emit-receive process, forces produced by an intensity-modulated optical signal in an emitter waveguide transduce a coherent travelling-wave phonon. The device structure is designed to shape the transfer function of this phononic signal to a nearby receiver waveguide, which converts the signal back to the optical domain as phase modulation through photoelastic coupling. The emit-receive process was used to create filter functions with Q-factors in excess of 1500 at GHz frequencies with tailorable, ultrasharp frequency rolloffs \protect{\cite{shinpper}}. Because this filtering process is mediated by GHz-frequency acoustic phonons, rather than light, this narrowband operation is supported without the use of high-Q optical resonators which experience nonlinear distortion at low optical powers. To adapt this nascent technology to RF-photonic filtering applications, however, it is necessary to understand the performance capabilities and limitations of such emit-receive processes within practical RF-photonic systems.

In this article, we report a new all-silicon photonic-phononic emitter-receiver (PPER) that produces record-high internal efficiency and use this device to demonstrate high-fidelity narrowband filtering within an RF-photonic link. We leverage a dual-core silicon waveguide that guides both light and sound to produce efficient photon-phonon coupling with an interaction strength sufficient to preserve or amplify modulation of optical signals through the PPER process. This structure produces a narrowband (5 MHz) filter response with high optical power handling ($>$100 mW) as a basis for low-loss integrated RF-photonic filters. Through systematic study of this device inside of an RF-photonic link, we demonstrate high-fidelity signal transfer with low insertion loss; a net RF link gain of $G$ = -2.3 dB and spur-free dynamic range SFDR$_3$ = 99 dB Hz$^{-2/3}$ are measured. We show that this architecture is uniquely suited for operations that require cascaded filter banks with minimal loss of signal fidelity. Building on these results, we discuss the potential for further improvements in the performance of RF-photonic PPER systems.


\section{Results}

\subsection{Operation Scheme}

The basic operation scheme of the photonic-phononic emitter-receiver system is diagrammed in Fig. 1a. The device consists of two Brillouin-active optical waveguides coupled to a common acoustic phonon mode. In typical operation, an intensity-modulated light signal (Fig. 1b.ii) is incident in port 1 of the device (labeled the `emit' waveguide). When the modulation frequency of this signal ($\Omega_\textup{s}$) is equal to the Brillouin frequency ($\Omega_\textup{B}$) of the device, the phonon is excited with signal frequency $\Omega_\textup{s}$ through a forward-SBS process (Fig. 1b.iii). This acoustic wave then phase-modulates a probe wave incident in port 2 of the device (the `receive' waveguide) through a linear acousto-optic scattering process (Fig. 1b.iv). Phase-matching permits phonon-mediated coupling for this process even when signal and probe waves are disparate in wavelength ($>$200 nm) \protect{\cite{shinpper}}. The phase-modulated signal then exits port 4 of the device, where it can be converted back to intensity modulated light through use of a frequency discriminator (e.g. interferometer, filter, phase-shifting element, etc. \protect{\cite{urick2015fundamentals}})  and the modulated light can then be used for RF-over-fiber, integrated signal processing, or converted into an RF signal via a photodiode. The frequency response of the PPER is determined by the engineered acoustic response of the system, which can be tailored for specific applications. Depending on the desired filter bandwidth, the acoustic phonon that mediates the emit-receive process can be designed to be travelling-wave, resonant, or a super-modal excitation of a collection of acoustic resonators \protect{\cite{shinpper}}. 


\begin{figure}[t]
\centering
\includegraphics[width=\linewidth]{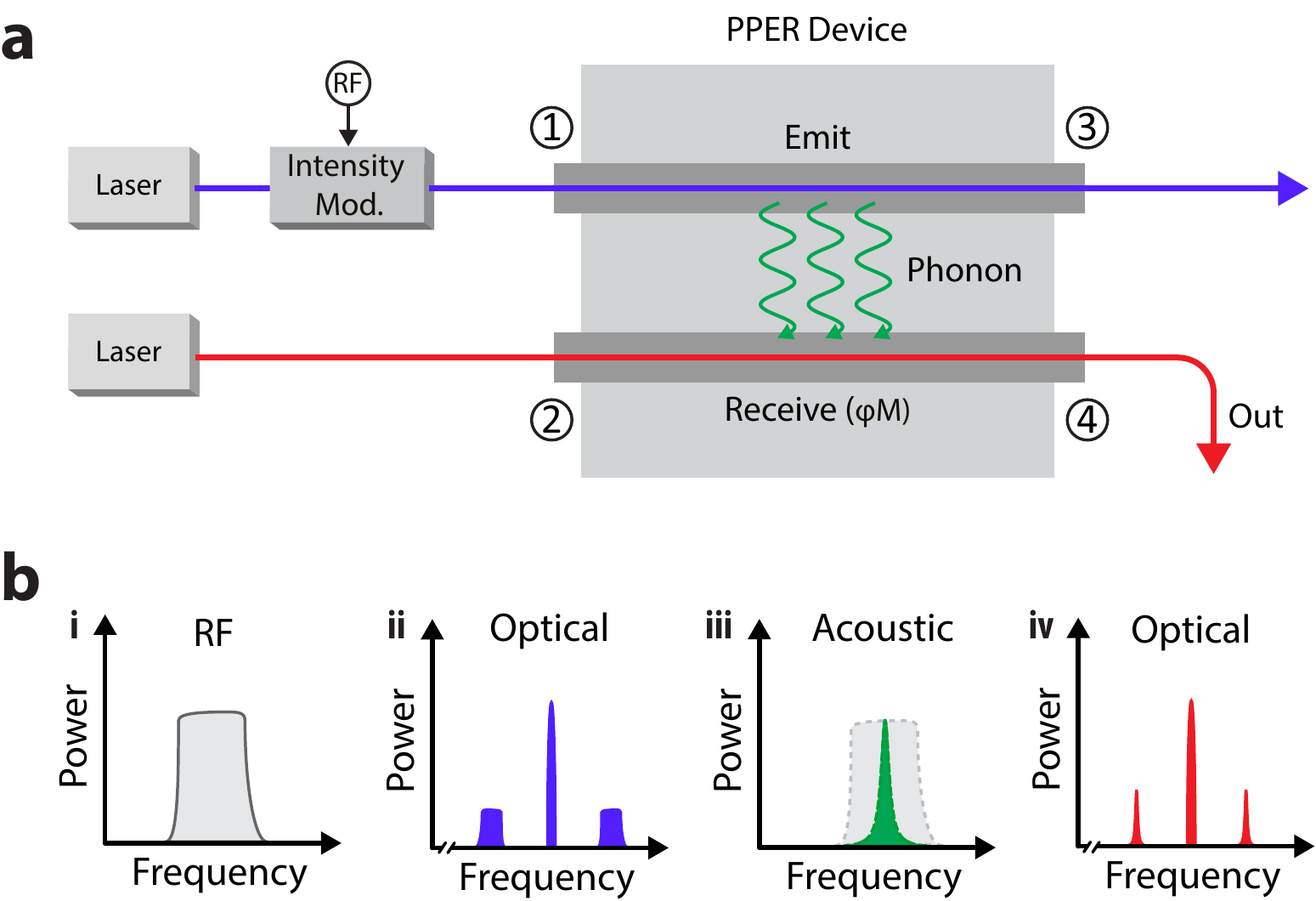}
\caption{Operation scheme of the photonic-phononic emitter-receiver. (a) depicts a diagrammatic representation of the PPER process. An optical signal wave (blue) is intensity modulated at microwave frequencies and injected into port 1 of the PPER device. The modulation is encoded on an acoustic phonon (green) which transduces information onto light injected into port 2 as phase modulation. Signal light exits port 3 of the device where it can then be used for further operations. Probe light exits port 4 of the device where it can be converted back into a microwave signal. (b) depicts (i) an example incident broadband RF spectrum (ii) the optical spectrum of the signal wave resulting from intensity modulation (iii) the acoustic power spectrum resulting from a combination of the original broadband RF modulation and the device's engineered acoustic transfer/filter function, and (iv) the filtered signal imparted on the probe beam as phase modulation.}
\label{fig:pper}
\end{figure}

In summary, the PPER can be understood as a device that (1) converts intensity modulation to phase modulation and (2) acts as an RF filter with a transfer function set by the device's acoustic response. At the same time, this process can be used to implement wavelength conversion due to the large phase-matching bandwidth of the receive-waveguide scattering process.


\subsection{Silicon Waveguide Photonic-Phononic Emitter-Receiver}

We study the photonic-phononic emit-receive operation using the optomechanical waveguide diagrammed in Fig. 2a. This PPER structure is fabricated from a single-crystal silicon layer using an SOI fabrication process. The device consists of a silicon membrane which guides optical waves in the fundamental optical modes of two separate ridge waveguides (Fig 2d-f). The emit-receive functionality in this structure is mediated by a resonant Lamb-like acoustic mode (Fig 2g) confined to the device by the large acoustic impedance mismatch between silicon and air. In this process, the acoustic wave is driven through a forward-SBS process in the emit waveguide. This optically-driven transverse acoustic wave time-modulates the effective index of the structure, imparting phase modulation on probe light in the receive waveguide. The core widths of these two waveguides are designed to be different by 100 nm. This changes the relative propagation constants of light to inhibit evanescent coupling between the emit and receive waveguides. As a result, the two channels are optically isolated and information is transferred only through phononic transduction.

\begin{figure*}[t]
\centering
\includegraphics[width=\linewidth]{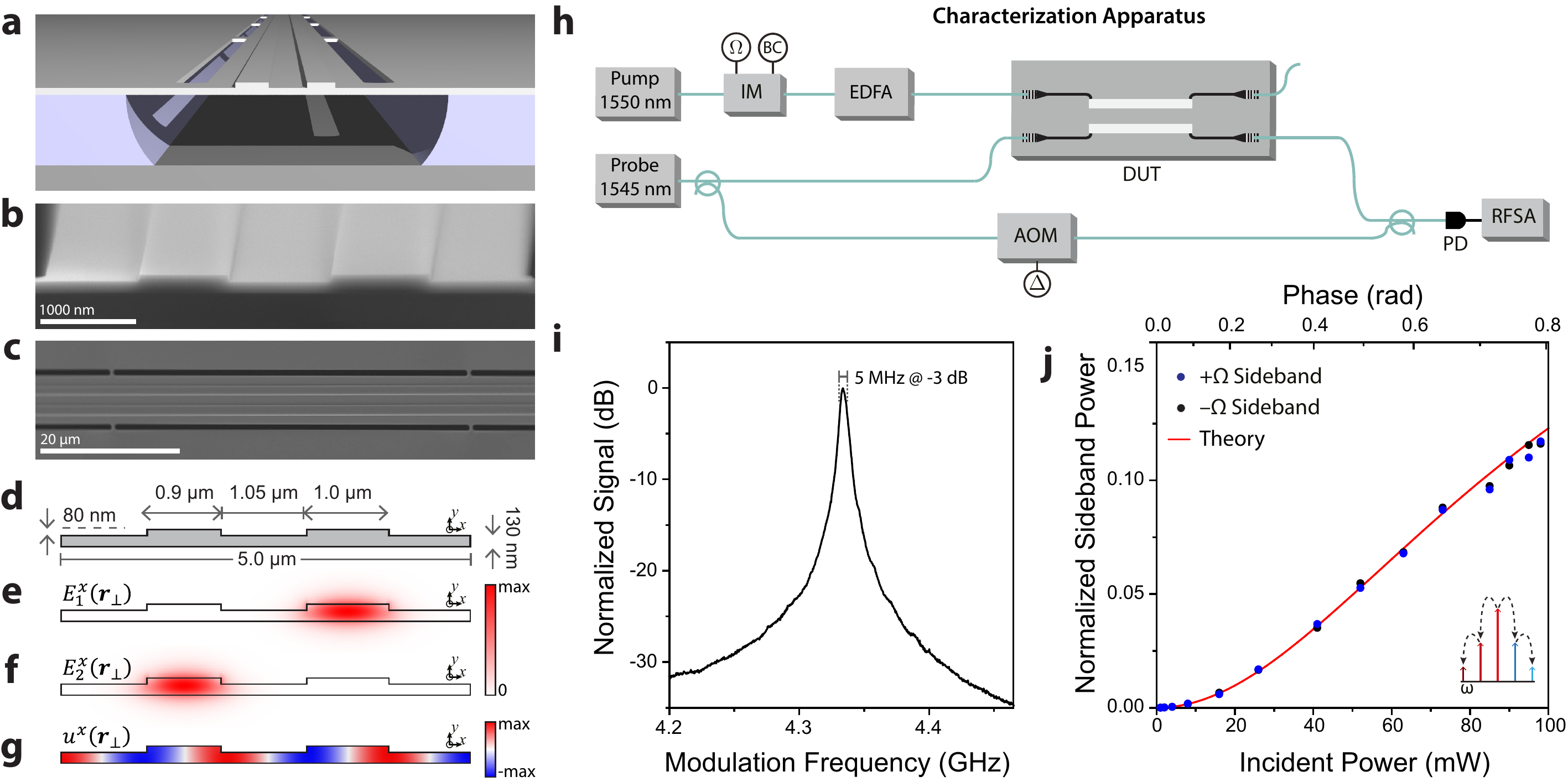}
\caption{Silicon photonic-phononic emitter-receiver. (a) Artistic representation of device; a dual-ridge structure is suspended by a series of nanoscale tethers. (b) scanning electron micrograph (SEM) of device cross-section. The scale bar represents 1000 nm in length. (c) Top-down SEM of suspended device, with a scale bar representing 20 $\upmu$m. (d) Diagram of device cross-section with dimensions listed. (e) and (f) are the $E_x$ fields of the two guided optical modes which couple through a $\sim$4.3 GHz Brillouin-active acoustic mode whose $x$-displacement field is plotted in (g). (h) depicts an experiment for measuring the PPER response. Pump light is intensity modulated around the zero-bias point and amplified to create two strong pump waves separated by a frequency around the Brillouin resonance frequency. Probe light is passed through the device and its output spectrum characterized by heterodyne detection using a frequency-shifted local oscillator. (i) plots the normalized frequency response of the device as a function of modulation frequency. (j) plots the sideband powers normalized to the incident probe wave power as a function of incident pump power.}
\label{fig:device}
\end{figure*}

To quantify the performance of this system, we begin by performing basic measurements of the modulation efficiency and frequency-dependent response of the PPER device. In this experiment, two strong optical tones of equal amplitude separated by around the Brillouin frequency of the device are injected into the emit waveguide. These tones drive a single-frequency acoustic field which imparts phase modulation on a probe beam incident in the receive waveguide, with a phase shift related to the total incident pump power. In the absence of optical losses, this phase shift is simply $\phi = G_{\textup{B}} (\Omega) P_{\textup{p}} L,$ where $G_{\textup{B}}(\Omega)$ is the frequency-dependent Brillouin gain coefficient, $P_{\textup{p}}$ is the power in each drive tone, and $L$ is the device length.

This experiment is carried out using the characterization apparatus diagrammed in Fig. 2h. Two strong pump waves with frequencies $\omega_{\textup{p}} \pm \Omega/2$ are synthesized by passing light from a pump laser through an intensity modulator operating in the null-bias regime. The amplitude of these waves is controlled using an erbium-doped fiber amplifier and variable attenuator. Pump light is coupled in and out of the emit waveguide of the PPER device using integrated grating couplers. A probe wave with frequency $\omega_{\textup{pr}}$ from a different laser is split into two paths; one is coupled on-chip while the other is passed through an acousto-optic modulator to produce a frequency-shifted local oscillator at frequency  $\omega_{\textup{p}} + \Delta$ = $\omega_{\textup{p}} + 2\pi\times44$ MHz. After probe light passes through the receive waveguide of the PPER device, it is recombined with the local oscillator arm to perform heterodyne spectral analysis for the transmitted spectrum in the microwave domain using a fast photodiode and radio-frequency spectrum analyzer.

\begin{figure*}[t]
\centering
\includegraphics[width=.8\linewidth]{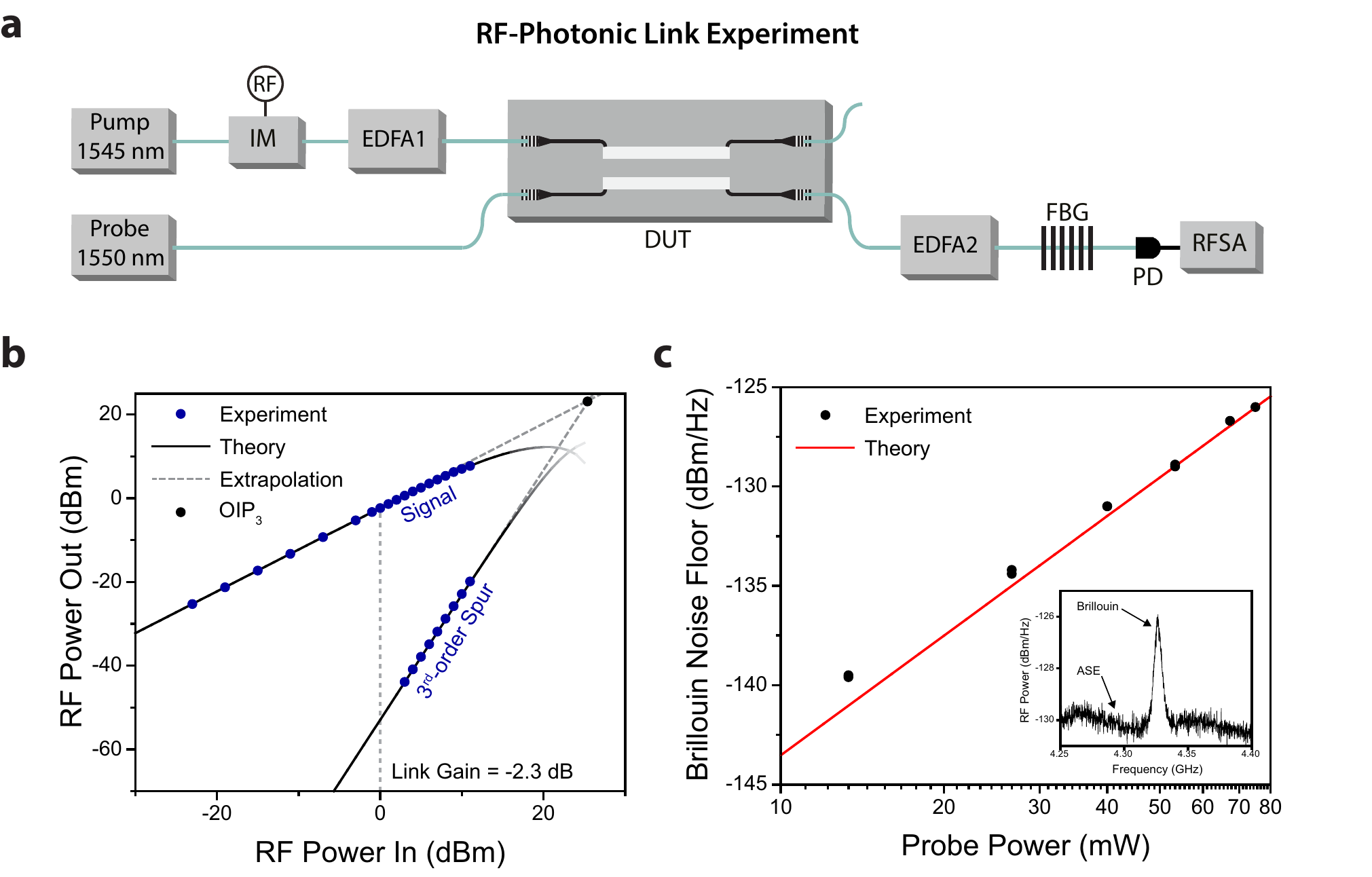}
\caption{RF-photonic link performance and noise analysis. (a) Diagram of the RF-photonic link with PPER device. Pump light is intensity modulated about the quadrature point by an input RF signal and amplified before passing into the emit port of the device. Probe light passes through the receive port and is amplified after coupling off-chip to compensate for coupling losses. A fiber Bragg grating is used as a narrowband filter for frequency discrimination, and a photodiode converts the signal back into the RF domain. (b) RF input/output data for the fundamental (top) and third harmonic (bottom) tones of the RF-photonic link. These are used to find the third-prder intercept point OIP$_3$ via extrapolation (dashed grey lines) (c) Characterization of the Brillouin noise floor at $\Omega/2\pi = 4.33$ GHz as a function of probe power. An inset shows the Brillouin noise spectrum and background due to amplified spontaneous emission from the EDFA.}
\label{fig:rfpl}
\end{figure*}

As the pump-wave detuning, and hence modulation frequency, is swept through the Brillouin frequency $\Omega = \Omega_{\textup{B}},$ efficient phase modulation of the probe wave is observed, resulting in optical energy transfer to frequency-detuned sidebands at  $\omega_{\textup{pr}} \pm \Omega$. Fig 2i plots the amplitude of the first phase-modulated sideband at frequency $\omega_{\textup{pr}} + \Omega$ as a function of input modulation frequency. These data reveal a narrowband (Q $\sim$ 820) resonance corresponding to a Brillouin-active phonon at $\Omega_{\textup{B}}/2\pi = 4.33$ GHz. As the total pump power is increased to a maximum of 100 mW, the relative sideband amplitude increases to a maximum of 12\%, corresponding to a phase shift of $0.8$ rad. These metrics represent a $10-$fold improvement in efficiency over previous devices of this type \protect{\cite{shinpper}}. While at low powers this efficiency follows the expected quadratic dependence ($\propto \phi^2$), at high powers it saturates due to a combination of the Bessel-like phase modulation response and nonlinear absorption in the emit waveguide. These data correspond to a Brillouin gain coefficient $G_{\textup{B}} = 820 \pm 20$ W$^{-1}$m$^{-1}$ for this $L = 2.53$ cm long PPER device. This effective nonlinear coefficient results from a combination of the Brillouin couplings in the emit and receive waveguides. In reality, these waveguides have different single-waveguide gain coefficients due to the engineered asymmetry of the PPER device. The individual gain coefficients for each waveguide were found through standard Brillouin gain measurements \protect{\cite{Kittlaus2016}} to be $G_{\textup{B}} = 880 \pm 50 $ W$^{-1}$m$^{-1}$ for the emit waveguide and $G_{\textup{B}} = 740 \pm 50 $ W$^{-1}$m$^{-1}$ in the receive waveguide. These values agree well with simulated values of $G_{\textup{B}} = 860 \pm 130 $ W$^{-1}$m$^{-1}$ and $G_{\textup{B}} = 790 \pm 120 $ W$^{-1}$m$^{-1},$ respectively.

\subsection{RF-Photonic Link Performance}

We next show that the performance of this device is sufficient to support efficient filter architectures in an RF-photonic link. First, we characterize the input/output RF response of this system. We then identify the dominant sources of distortion and noise through this link.

We explore RF-photonic performance of the PPER device using the RF-photonic link diagrammed in Fig 3a. An RF signal is encoded on a pump beam using a commercial intensity modulator (Optilab IM-1550-20, $V_\pi = 9.9$ V) biased at the quadrature point. Light is amplified using an erbium-doped fiber amplifier and coupled into the emit waveguide using an integrated grating coupler. Probe light from a separate laser is coupled into the receive waveguide, where the signal is filtered through the PPER device's phononic response and encoded onto the probe wave as phase modulation. After passing through the active device region, this light is coupled off-chip, amplified through an erbium-doped fiber amplifier to offset fiber-chip coupling losses, and converted to intensity modulation by filtering out one of the $\pm \Omega$ modulation sidebands. This signal is then converted back to the RF-domain using a commercial high-power photodiode (Discovery Semiconductors, Inc. DSC100S, responsivity $R = 0.75$ A/W, bias voltage $V_{\textup{b}} = 7$ V). 

Link gain is measured using a single RF input tone at $\Omega = \Omega_{\textup{B}}/2\pi = 4.33$ GHz and with incident on-chip optical pump power $P_p = 105$ mW and probe power at detector $P_{\textup{det}} = 75$  mW. The corresponding RF signal input/output curve is plotted as the top curve in Fig. 3b. With these specifications, the net RF-photonic link gain is -2.3 dB. The 3 dB bandwidth of the RF filter response is identical to the device's phononic bandwidth of $5 $ MHz measured in Fig. 2i. 

Independent of the external RF components, the performance of the PPER device can be understood by comparing the output phase modulation strength to the input intensity modulation strength. Through these experiments, the effective phase modulation produced through the PPER device is $\phi_{\textup{PM}}^{\textup{out}} = 0.559$ rad/V, compared to an input intensity modulation of $\phi_{\textup{IM}}^{\textup{in}} = 0.317$ rad/V. We define a figure of merit which is the ratio between these two modulation strengths which can be applied to any modulation-conversion operations of this type. In this system, $\phi_{\textup{PM}}^{\textup{out}}/\phi_{\textup{IM}}^{\textup{in}} = 1.76,$ indicating a net enhancement in modulation through the silicon PPER device. 


We next seek to quantify the dominant forms of distortion in the PPER RF-photonic link. The first form of distortion in RF-photonic links is compression of the linear response at high input RF powers. This is quantified through measuring the RF output power at 1 dB compression from a linear response. The power output at 1 dB compression P$_{1\textup{dB}} = 7.7$ dBm through the PPER filter. 

The second form of distortion we consider is that due to third-order spurious tones generated through the RF-photonic link. Due to the narrowband filter response of the PPER system, out-of-band frequency components are rejected by the phononic filter. Therefore spurious tones must be either (1) generated at harmonics of $\Omega_{\textup{B}}$ due to phonon-mediated phase modulation in the receive waveguide or (2) unwanted frequency components around $\Omega_{\textup{B}}$ created by the intensity modulator at the frontend of the link. For most systems and device parameters, the latter is the dominant source of spurious tones, as is the case here. Through this process, frequency mixing during modulation between input frequencies not within the PPER filter bandwidth can result in new frequency components that can pass through the filter. We quantify the strength of this effect by driving the system at a frequency $\Omega = \Omega_{\textup{B}}/3$ and examining the strength of the third-order spurious signal passing through the PPER at $\Omega = \Omega_{\textup{B}},$ plotted as the bottom curve in Fig. 3b. These spurs correspond to an output third-order intercept point OIP$_3$ of 23 dBm through the PPER system (black circle in Fig. 3b).


To characterize the dynamic range of this RF system, we finally carry out careful measurements of the sources of noise through the PPER device. The frequency-dependent noise floor around the Brillouin frequency is plotted as an inset in Fig. 3c. At the filter frequency ($\Omega_\textup{B}/2\pi = 4.33 $ GHz), the noise spectrum is dominated by spontaneous scattering from thermal phonons (population at room temperature $k_{\textup{B}} T / \hbar \Omega_{\textup{B}} \approx 1390)$ \protect{\cite{kharelpra,vanlaernoise}}, with an out-of band noise floor interpreted as amplified spontaneous emission from the output EDFA (labeled EDFA2 in Fig. 3a) in the link. The probe power dependence of the Brillouin noise is quantified in Fig 3c. For detector powers of 75 mW used in this measurement, the Brillouin noise power spectral density $N =$ -126 dBm/Hz. This noise power corresponds to a receive waveguide Brillouin gain coefficient of $G_{\textup{B}} = 730 \pm 40 $ W$^{-1}$m$^{-1}$, in good agreement with the numbers obtained from stimulated gain measurements. This fundamental noise source can be mitigated by designing PPER devices which couple through higher-frequency phonons with lower thermal occupations at room temperature.

Combining the measurements of total system nonlinearities and noise, we find a spur-free dynamic range SFDR$_3 = 2($OIP$_3$ - $N)/3 = 99.3 $ dB Hz$^{-2/3}$  normalized to a 1 MHz bandwidth. This corresponds to the dynamic range over which a signal is detectable through the link while third-order spurs are not. The linear dynamic range is CDR$_{1\textup{dB}} = $ P$_{1\textup{dB}} - N = $ 135 dB Hz$^{-1}$. In the next section we discuss use cases for the narrowband PPER filter as well as potential improvements to device performance.

\section{Discussion}

We have characterized the performance of the integrated PPER as both an active optical device and as a narrowband RF-photonic filter. This system supports a record-high nonlinear figure of merit for emit-receive systems of this type in an all-silicon device structure. Next, we discuss the potential for further improvement of the system as well as architectures for scalability and integration.

We have shown that new photonic-phononic emit-receive operations provide a promising path towards high-performance narrowband RF-photonic filters. Looking forward, several potential refinements of the PPER link and device are possible to produce higher performance in a significantly smaller device footprint. First, the performance of the PPER filter system can be readily enhanced by improving the modulation and detection schemes external to the filter. The intensity modulator used in the frontend of the link demonstrated here has a relatively high half-wave voltage $V_\pi = 9.9 $ V. Decreasing $V_\pi$ would allow the reduction of device length by a corresponding factor without affecting link gain or dynamic range. In addition to shrinking the device footprint, this reduced length would lead to a lower spontaneous noise floor, allowing the PPER system to operate on smaller RF signals as may be desirable in practical applications. Many modulator technologies have been already demonstrated in silicon with $V_\pi < 1 $ V \protect{\cite{Campenhout09,VanCamp12}}, demonstrating the feasibility of this approach. In addition, the current method used to convert phase to intensity modulation is sub-optimal since filtering out one of the two signal sidebands discards half of the signal power. The link gain of the system could therefore be improved by replacing this filter with interferometric balanced detection or a phase-shifting element.

Significant improvements in device performance can also be achieved by increasing optical power in the emit waveguide, which directly enhances the modulation efficiency of the PPER device. The pump power in the current device was limited to $\sim100$ mW due to nonlinear absorption  \protect{\cite{Kittlaus2016}}. This can be mitigated through electrical removal of free carriers generated due to two-photon absorption (TPA) \protect{\cite{rong2005}} or by changing the pump wavelength to be above the TPA threshold for silicon ($\lambda_\textup{p} > 2.2$ $\upmu$m). These would allow pump powers $>$2 W to propagate through the emit waveguide with negligible nonlinear losses, allowing dramatic reduction in device length and probe power to enable lower-noise operation with 5-10 dB improvements in dynamic range. Shorter devices are also known to have a smaller degree of inhomogeneous broadening of the Brillouin resonance \protect{\cite{wolffbroad}}. Therefore shorter device lengths would decrease the filter width and increase device performance, which depends directly on acoustic quality factor. A modified silicon PPER system with the parameters ($P_{\textup{p}} = 2$ W, $\lambda_{\textup{p}} = 2.75 $ $\upmu$m, $P_{\textup{pr}} = 10$ mW, $\lambda_{\textup{pr}} = 1.55 $ $\upmu$m, $V_\pi = 1$ V, $Q = 1500, L = 200$ $ \upmu$m) would have superior linearity (link gain $G$ = 0 dB, SFDR$_3$ = 105 dB  Hz$^{-2/3}$, CDR$_{1\textup{dB}}$ = 146 dB Hz$^{-1}$) in a much smaller footprint. The dynamic range of this system could be further improved by using a linearized modulation scheme, which have been studied extensively for RF-photonic applications \protect{\cite{Johnson88,Betts94,Bridges95}}.    

Unlike many conventional RF filters, the PPER filter allows for serial cascading without significant signal degradation, making it particularly attractive for signal channelization. Once an RF signal is encoded on the pump wave, it can pass through several successive PPER devices to select out separate spectral channels without impacting the fidelity of the original signal, as diagrammed in Fig 4a. By contrast, traditional channelization techniques based on band-pass filters require power to be split evenly between each channel (Fig 4b), resulting in noise increases due to prior amplification stages that grow with the number of channels. As an example application, a cascaded array of 200 PPER devices with the improved parameters discussed above could span a 500 MHz bandwidth with 2.5 MHz bandwidth per filter channel, while maintaining the same dynamic range and sensitivity as a basis for real-time spectral analysis.

\begin{figure}[t]
\centering
\includegraphics[width=\linewidth]{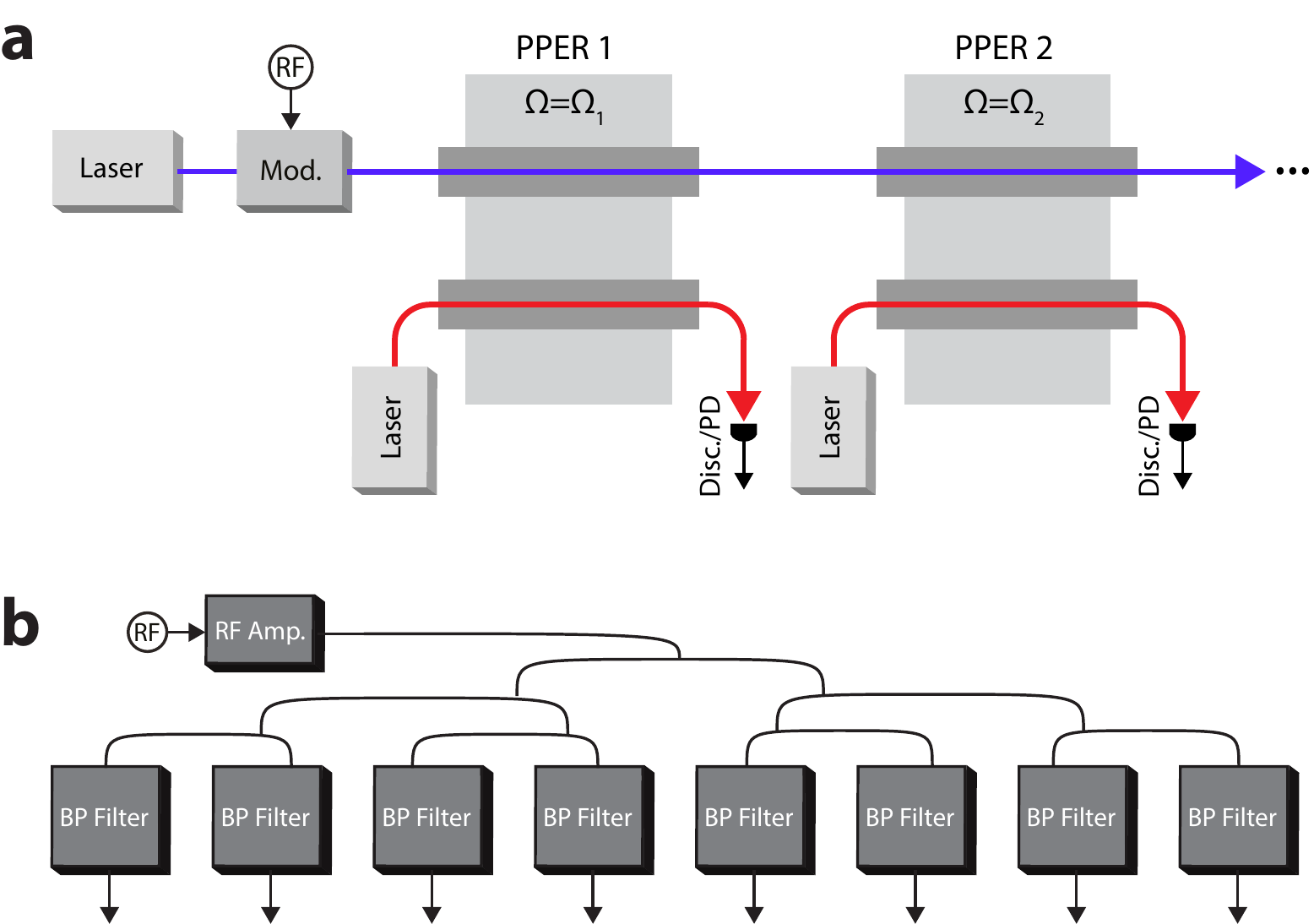}
\caption{Comparison of serial and parallel filter architectures for channelization. (a) Serial cascaded architecture for a PPER channelization filter. Because the same pump light can be used for each device (i.e. because the PPER device makes a copy of a narrowband slice of the input RF spectrum), minimal signal degradation is experienced as the number of channels is increased. (b) Architecture based on parallel RF filters. Note that because the signal must be split equally among the channels, amplification is necessary to preserve signal amplitude.}
\label{fig:casc}
\end{figure}

Emit-receive operations can be realized in many systems where optical fields guided in distinct modes or polarizations couple through the same phonon mode, provided that linear and nonlinear optical crosstalk between channels is negligible. Current and potential avenues of research include multi-core fibers \protect{\cite{HagaiDiamandi17}}, optomechanical waveguides coupled by acoustic supermodes \protect{\cite{shinpper}}, and light fields of disparate wavelengths \protect{\cite{shinnatcomm,vanlaernatphoton}} or spatial modes \protect{\cite{Kittlaus2017}} within the same Brillouin-active waveguide core. Building on the present work, PPER interactions may be harnessed in a variety of systems for applications in optical and RF signal processing and spectral analysis.

The integrated silicon-based PPER device demonstrated here can be readily interfaced with existing on-chip technologies for silicon-based modulators \protect{\cite{Xu2005, Reed2010}} and integrated photodetectors \protect{\cite{Chen09,tbj2012,Xie16}}. Such on-chip integration would allow low-cost implementation of such filters in complex RF-photonic circuits. Like other acousto-optic based RF-photonic filters, the PPER can also be used for frequency-agile operation; while the PPER device implemented in the RF-photonic link here operates at a fixed frequency, it is straightforward to adapt Brillouin-based RF-photonic filter systems to frequency agile operation by changing the modulation scheme \protect{\cite{marpaung13,marpaung,shinpper,CasasBedoya15}}. 



In conclusion, we have demonstrated an all-silicon hybrid photonic-phononic filter and characterized its performance within an RF-photonic link. This device supports record-high modulation efficiency for phononic emit-receive devices and robust performance as an RF-photonic filter. This work demonstrates the potential for large-scale integration of these agile filters into RF-photonic interconnects with applications in communications, sensing, and signal analysis.

\section{Methods}

\subsection{Device Fabrication and Optical Performance}
The silicon waveguides were written on a silicon-on-insulator chip with a 3 $\upmu$m oxide layer using electron beam lithography on hydrogen silsesquioxane photoresist. After development, a Cl$_2$ reactive ion etch (RIE) was employed to etch the ridge waveguides and integrated grating couplers. After a solvent cleaning step, slots were written to expose the oxide layer with electron beam lithography of ZEP520A photoresist and Cl$_2$ RIE. The device was then wet-released in 49\% hydrofluoric acid to remove the oxide undercladding. The device under test is comprised of 498 suspended segments.

These waveguides have an estimated linear optical propagation loss of $0.2$ dB/cm from cutback measurements. The nonlinear absorption coefficients in units of guided-wave power are $\beta^{\textup{TPA}} \approx 50$ W$^{-1}$m$^{-1}$ for two-photon absorption and $\gamma^{\textup{FCA}} \approx 2400$ W$^{-2}$m$^{-1}$ for two-photon absorption-induced free-carrier absorption as measured through nonlinear power transmssion experiments.

\subsection{Experiment}
The following abbreviations are used in the experimental diagrams: IM Mach-Zehnder intensity modulator, BC bias controller, EDFA erbium-doped fiber amplifier, DUT device under test, AOM acousto-optic frequency shifter, FBG fiber Bragg grating, PD photodetector, RFSA radio-frequency spectrum analyzer.  




\bibliographystyle{naturemag-ed} 
\bibliography{cites}





\end{document}